%% file: p23.tex
\documentclass[12pt,reqno]{amsart}
\usepackage{graphicx}
\usepackage{amssymb,amscd,amsbsy}
\usepackage{amsthm}
\input{def23}

\theoremstyle{remark}
\newtheorem{rem}[thm]{Remark}
\newtheorem*{rem*}{Remark}

\begin{document}

\newcommand{\vse}{\vspace{.2in}}
\numberwithin{equation}{section}

\title{\bf THE ATIYAH-HITCHIN BRACKET FOR THE CUBIC NONLINEAR SCHR\"{O}DINGER EQUATION.\newline I.  GENERAL POTENTIALS.}
\author{K.L.  Vaninsky}
\thanks{ The work is partially supported by NSF grant DMS-9971834.}
\begin{abstract}
This is the first in a series of  papers on Poisson formalism  for the cubic nonlinear Schr\"{o}dinger 
equation with repulsive nonlinearity and its relation to  complex geometry.  In this paper we study general continuous potentials.  
We demonstrate that  the Weyl functions of the corresponding  auxiliary Dirac spectral  problem carry a natural Poisson structure. We call it the Atiyah--Hitchin Poisson bracket. We show that the Poisson bracket on the 
phase space is the image of the Atiyah--Hitchin bracket on Weyl functions 
under the inverse spectral transform. 
   
\end{abstract}
\maketitle

\setcounter{section}{0}
\setcounter{equation}{0}
\section{Introduction.}
\subsection{Statement of the problem.}
 All 1+1  differential equations like the Korteweg-de-Vries, the modified  Korteweg-de-Vries, 
the cubic  nonlinear Schr\"{o}dinger equation, the Toda lattice, the Camassa--Holm equation {\it etc.,} which are   analyzed by the
inverse spectral transform, are  Hamiltonian systems. 

The cubic NLS with repulsive nonlinearity\footnote{Prime $'$  signifies the derivative in the   variable $x$ and dot $\bullet$ the derivative with 
respect to time.}  
$$
i \psi^\bullet= - \psi'' + 2 |\psi |^2 \psi,
$$
where $\psi(x,t)$ is a complex function, 
will serve as our model example. We consider this problem on the entire line, {\it i.e.}, $x \in \RB^1$. 
We assume that the phase space $\MM$ consists of  functions $\psi(x)$ and we do not impose any condition on $\psi(x)$ except continuity. We restrict our attention to some specific function classes, {\it e.g.},  
periodic or  rapidly decaying, in subsequent  papers. 
To make apparent the algebraic nature of our considerations,   we assume that all functionals $A,B: \MM \rightarrow \C$ are Frechet differentiable, all integrals converge, {\it etc.} 

The cubic NLS  equation is  a Hamiltonian system 
$$
\psi^{\bullet}= \{\psi, \HH\},
$$
with  the classical bracket
\beq\label{cb}
\{A,B\}=2i \int {\d A\over \d \psib(x)}{\d B\over
\d \psi(x)}- {\d A\over \d \psi(x)}{\d B\over 
\d \psib(x)}\, dx, 
\eeq
and   Hamiltonian 
$$\HH={1\over 2} \int |\psi'|^2 
+|\psi|^4 \, dx. 
$$ 

The NLS equation arises as a compatibility condition for  the commutator relation for some specially chosen differential operators. This  leads to 
an auxiliary linear spectral problem for the  Dirac operator
$$
{\Dg} \f= \[\(\begin{array}{ccccc} 1& 0 \\ 0 &  -1  \end{array}\) i\partial_x +
\(\begin{array}{ccccc}
 0& -i \overline{\psi} \\
 i \psi & 0 \end{array}\)\] \f= \lt \f
$$
acting in a space of vector functions 
$$
\f(x,\l)=\[  \begin{array}{ccccc}
                 f_1(x,\l)\\
                 f_2(x,\l) \end{array}\].  
$$
The goal of this  and subsequent papers is to demonstrate that the Hamiltonian formalism (Poisson bracket)  is  built into the complex geometry of the Dirac  spectral problem.  

\subsection{Description of results.}

We associate to a potential $\psi(x)$ of the  Dirac spectral problem a pair $(\G,\P)$, {\it i.e.}, we 
have a map:
$$
\MM \quad \longrightarrow \quad (\G,\P). 
$$
We call this map the direct spectral transform. 
In this pair  $\G$  is a two sheeted 
covering of the complex plane of the  spectral parameter  cut along the real line (see Figure 1). 
\begin{figure}[htb]
\includegraphics[width=0.60\textwidth]{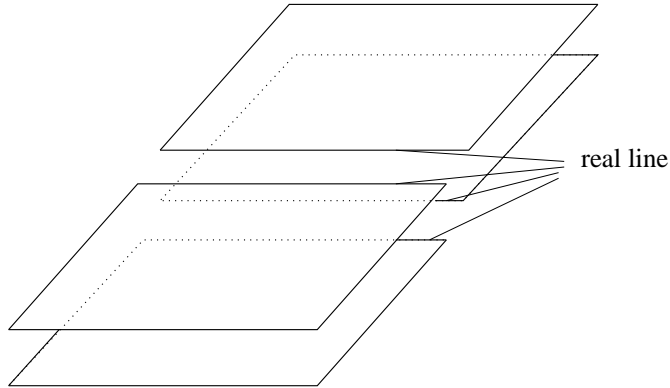}
\caption{ The spectral cover.}
\end{figure} 
\noi
For any point $Q\in \G$, its projection to the 
plane of the spectral parameter $\l$  is denoted by $\l(Q)$. The holomorphic  function $\P(Q)$ defined on $\Gamma$ is the so-called Weyl function of the  Dirac spectral problem.    We show that  the pair $(\G,\P)$  carries a natural Poisson structure. We call it the Atiyah-Hitchin bracket. Apart from insignificant constants 
the AH bracket is given by the formula
\beq\label{ah}
\{\P(Q),\P(P)\}={(\P(Q)-\P(P))^2 \over \l(Q)- \l(P)}. 
\eeq
This is the first theorem  of the paper. 

It turns out that the direct spectral transform  can be inverted 
$$
(\G,\P)\quad \longrightarrow \quad \MM .
$$
We call this map the inverse spectral transform. 
We show that  the image of the AH bracket   
under the inverse spectral transform becomes the  bracket \ref{cb}  on  phase space. 
This is our second  theorem.

The pair $(\G,\P)$ introduced in this paper is a skeleton of an analogous   construction for  periodic and scattering potentials in subsequent 
papers \cite{V4} and \cite{V5}. In these cases 
there is an additional structure which allows one to identify the values of the meromorphic function $\P$ on different banks of the cut.

The question of  construction of canonical coordinates can not be resolved without assumptions on a potential. Canonical coordinates are given  in terms of singularities of the Weyl functions on the real line, or in other words in terms of the spectrum of the auxiliary operator.  
At the moment the only equations  which have been analyzed
are  the finite Toda lattice, \cite{V1}, and the Camassa--Holm equation 
with rapidly decaying  initial data of one sign, \cite{V2}. For both of these systems the spectrum of the corresponding auxiliary spectral problem is an isolated discrete set\footnote{For the Camassa-Holm equation  and the open Toda lattice the corresponding Poisson bracket on the  Weyl function is  also given  by formula \ref{ah} or its' reduction.} . 
 We refer to \cite{V1} and 
\cite{V2} for details.

\subsection{Historical remark.} 
Another  approach which  relates   Hamiltonian theory and complex geometry  was 
developed  recently.

In a  remarkable  paper Krichever and Phong \cite{KP}
proposed a  construction of a {\it symplectic formalism}   for
integrable equations with  periodic initial data. For the latest exposition of their results and connections 
with  Seiberg-Witten theory see \cite{DKP}. 
The Krichever-Phong approach is designed for 2+1 systems like the Kadomtsev-Petviashvili equation  and the 2D-Toda lattice.
It can also be extended to the scattering case, \cite{V3}. In the simplest cases the simplectic form can be inverted explicitly to obtain the 
Poisson bracket. For example the Gardner-Zakharov-Faddeev bracket for KdV or classical bracket \ref{cb} 
(both are constant on the phase space)  can be obtained this way. In the case 
of a bracket with variable coefficients the situation is different. It is not possible to invert explicitly the corresponding   higher (second) 
symplectic form even for the case of NLS with periodic boundary conditions, see \cite{V3}.    These prompt us  to find another approach to 
the Poisson formalism.

\subsection{Content of the paper.} In  Section 2 we define the Atiyah-Hitchin bracket  for  rational maps of $\CP^1$ into itself. We 
state its basic properties and prove that the AH bracket is invariant under linear fractional transformations.   The direct spectral transform and inverse spectral transform are defined in Section 3. We study the Poisson bracket in Section 4, which consists of two parts. In the first part we compute the image of the classical bracket for  the Weyl function. In the second part we show that 
the bracket on  phase space is an image of the AH bracket under the inverse spectral transform.

\section{The Atiyah--Hitchin bracket.} 
Let $\P(\l): \CP^1 \rightarrow \CP^1$  be the  rational function  of degree $N$  such that 
\begin{itemize}
\item  $\P(\infty)=0$, 
\item the preimage $\P^{-1}(\infty)$  consists of  exactly $N$ points.  
\end{itemize} 
To any $N$--monopole solution $m_N(x),\; x\in \RB^3,$ of the Bogomolny equation   
Atiyah and Hitchin, see \cite{AH},   
associate the {\it scattering function} $\P(\l)$, with these  properties. 
Thus we have  the direct spectral transform
$$
m_N(x)\quad \longrightarrow \quad \P(\l).
$$
This map is injective due to a theorem of S. Donaldson, \cite{AH}. 

The function $\P(\l)$ has the form
$$
\P(\l)={\sum\limits_{i=1}^{N-1} a_i \l_i\over \l^N + \sum\limits_{j=0}^{N-1}
b_j \l^j}=-{q(\l)\over p(\l)}.
$$
  The monic polynomial $p(\l)$ is determined by its roots $\l_1,\hdots,\l_{N}$. 
The polynomial $q(\l)$ of degree $N-1$ can be determined from its values at 
the roots of denominator.  Therefore,  
$$
\l_1,\hdots,\l_{N}, q(\l_1),\hdots,q(\l_{N});
$$
are global complex coordinates on this space of maps. 

Let $\d$ denote a variation of the parameters $\l_1,\hdots,\l_{N}, q(\l_1),\hdots,q(\l_{N})$, while 
$d$ is a differential  of the parameter $\l$.
The Atiyah--Hitchin symplectic structure 
$\omega$ is defined by the formula
$$
\omega=\sum_{k=1}^{N} \frac{\d q(\l_k)}{q(\l_k)} \wedge \d \l_k.
$$
The corresponding Poisson bracket is specified  by canonical relations:
\beq\label{ahcoo}
\{q(\l_n), \l_k\}=\d^n_k \; q(\l_n).
\eeq
All other brackets vanish
\beq\label{ahco}
\{\l_n, \l_k\}=\{q(\l_n), q(\l_k)\}=0.
\eeq
The bracket turns  the space of maps $\P(\l): \CP^1 \rightarrow \CP^1$ of degree $N$ with the above mentioned properties into a Poisson manifold. 

Consider some  function $\P_0$,  a point of the Poisson manifold. Fix some points $\l$ and $\m$ on the sphere away from the poles of  $\P_0$.
Then  the points $\l$ and $\m$ considered as an argument of  $\P$, where $\P$ is  from  a small vicinity of $\P_0$,  are well defined functions in this 
vicinity\footnote{Instead of $\P(\l)$ and $\P(\m)$ one should write $\l(\P)$ and $\m(\P)$.}. 
Their values at $\P$ are functions of the 
coordinates  $\l_1,\hdots,\l_{N}, q(\l_1),\hdots,q(\l_{N})$. 
As it was  demonstrated by Faybusovich and Gekhtman, \cite{FG}, the bracket for $\P(\l)$ and $\P(\m)$  is given by the formula
\beq\label{AH}
\{\P(\l),\P(\m)\}=\frac {(\P(\l)-\P(\m))^2}{\l-\m}.
\eeq
Here some miracle occurs. The Poisson bracket for two functions ($\l$ and $\mu$)  on the Poisson manifold  is given in 
terms  of these  functions  and   their argument ($\P$). 
In \cite{V1} for rational functions we gave a direct proof that   \ref{AH} implies 
\ref{ahcoo}--\ref{ahco}.  
We list    properties of the AH bracket. 

Evidently, \ref{AH} is skew--symmetric with respect to $\l$ and $\m$.
It is also linear in its arguments
\beq\label{linear}
\{a\P(\l)+b \P(\m),\P(\nu)\}=a\{\P(\l),\P(\nu)\} +b\{\P(\m),\P(\nu)\}, 
\eeq
where $a$ and $b$ are constants.  The Leibnitz rule  holds 
\beq\label{leib}
\{\P(\l)\P(\m),\P(\nu)\}=\P(\l)\{\P(\m),\P(\nu)\} + \P(\m)\{\P(\l),\P(\nu)\}. 
\eeq
The bracket  satisfies the Jacobi identity
$$
\{\P(\l),\{\P(\mu),\P(\nu)\}\}+\{\P(\m),\{\P(\nu),\P(\l)\}\}+
\{\P(\nu),\{\P(\l),\P(\mu)\}\}=0.
$$

The bracket is  compatible with the Cauchy theorem in the following sense. 
Any  function  $\P(\l)$ analytic  inside some  contour $C$ is determined by  
its the values at the contour
$$
\P(\l)={1\over 2\pi i} \int\limits_{C} {\P(\zeta)\over \zeta-\l} d \zeta, 
$$
for  $\l$ inside $C$. 
Whence  due to \ref{linear} the values of the bracket at different points are
related:  
$$
\{\P(\l),\P(\m)\}={1\over 2\pi i} \int\limits_{C} {\{\P(\zeta),\P(\m)\}\over
\zeta-\l} d\zeta, 
$$
for any $\mu \in \CP^1$. 
It can be verified directly  that the formula on the right in \ref{AH}   satisfies this 
compatibility condition.

The following is particularly useful to us.   
\begin{lem}\label{invan}
 The bracket \ref{AH} is invariant under linear fractional
transformations  
\beq\label{lft}
\P\qquad \longrightarrow \qquad \P'={a\P+ b\over c\P + d},  
\eeq
where $a,b,c,d$ are constants.
\end{lem}
 
{\it Proof.} Consider a transformation of the form
\beq\label{olft}
\P\quad \rightarrow \quad \P'={1\over c\P +d}.   
\eeq
Then
\bey
\{\P'(\l),\P'(\m)\}&=&{1\over (c\P(\l) +d)^2} {1\over (c \P(\m) +d)^2}
                              \{ c\P(\l) +d,c\P(\m)+d\}\\
&=& {1\over (c\P(\l)+d)^2}{1\over (c\P(\m)+d)^2} {((c\P(\l)+d) -(c\P(\m)+d))^2
\over \l -\m}\\ 
&=&{(\P'(\l)-\P'(\m))^2\over \l-\m}.
\eey
To finish the proof we note that two consecutive transformations of the form
\ref{olft}  produce the whole group \ref{lft}. \qed

The Atiyah--Hitchin bracket in  coordinate free form \ref{AH}
will appear  for a much wider class  than rational functions. 
Formula \ref{AH} itself can be a starting point for 
construction of the bracket on an infinite algebra of complex observables $\P(\l), \, \l \in \CP^1$, meromorphicaly dependent on the parameter. 
Indeed, we start defining the bracket for two observables at  different points by \ref{AH} and the extend it to all polynomials 
using \ref{linear} and \ref{leib}. 
It can be verified in a long but simple calculation that \ref{linear} and \ref{leib} 
imply the Jacobi identity for the AH bracket. It is of  interest to find all  formulas  similar to \ref{AH} with the  properties of a Poisson bracket and compatible with the Cauchy theorem. 

\section{The spectral problem.}

The  NLS equation, 
$$
i \psi^{\bullet}= -\psi'' + 2 |\psi |^2 \psi, 
$$
where $\psi(x,t)$ is  a  smooth complex function,   
is  a Hamiltonian system
$$
\psi^{\bullet}= \{\psi, \HH\},
$$
with  Hamiltonian
$\HH=\frac{1}{ 2} \int |\psi'|^2 +|\psi|^4 \, dx=${\it energy} and the bracket
\beq\label{nlspb}
\{A,B\}=2i \int \frac{\d A}{ \d \psib(x)}\frac{\d B}{ \d
\psi(x)}- \frac{\d A}{ \d \psi(x)}\frac{\d B}{ \d \psib(x)}\, dx.
\eeq

The  NLS equation is a compatibility condition for the commutator
$$
[\partial_t- V_3,  \partial_x-V_2]=0, 
$$
with\footnote{
Here and below  $\sigma$  denotes the {\it Pauli matrices}
$$
\sigma_1=\left(\begin{array}{ccccc}
 0& 1\\
1 & 0
\end{array}  \right),  \nonumber \quad
\sigma_2=\left(\begin{array}{ccccc}
0 & -i\\
i &  0
\end{array} \right), \nonumber \quad
\sigma_3=\left(\begin{array}{ccccc}
1 & 0\\
0 & -1
\end{array} \right). \nonumber
$$
}
$$
V_2 = - \frac{i \l }{ 2} \sigma_3 +Y_0 =
\(\begin{array}{ccc}
 - \frac{i\l}{ 2} &  0\\
 0 & \frac{i\l}{ 2} \end{array}\)  +
\(\begin{array}{ccccc}
  0& \overline{\psi} \\
 \psi & 0 \end{array}\)
$$
and
$$
V_3 = \frac{\l^2}{ 2}i \sigma_3 -\l Y_0 + |\psi|^2 i\sigma_3 -i \sigma_3 Y_0'.
$$
We often omit the lower index and write $V=V_2$.

\subsection{The direct spectral transform.} The commutator relation produces   an auxiliary linear problem  
\beq\label{spro}
\f'(x,\l)=V(x,\l) \f(x,\l), \qquad\qquad\qquad \f(x,\l)=\[  \begin{array}{ccccc}
                 f_1(x,\l)\\
                 f_2(x,\l) \end{array}\].  
\eeq
This can be written as an eigenvalue problem for the Dirac operator (see the introduction). 
Let  $\f^T$ denote the transposition of the vector $\f$ and let $\f^*$ denote the adjoint of the vector $\f$. Let $\LLL^2(a,b)$ be a space of vector functions with the property
$$
\int\limits_{a}^{b} \f^*(x,\l)\f(x,\l) \,dx < \infty.
$$
The {\it Weyl solution}, \cite{W},
  $$\e(x,y,\l)=\[\begin{matrix}
e_1(x,y,\l)\\
e_2(x,y,\l)
\end{matrix} \]$$
is the solution of \ref{spro}  which
belongs to  $\LLL^2[y,+\infty)$ or $\LLL^2(-\infty,y]$, where $y$ is an arbitrary point on the line. Due to the theorem of Levitan and Martunov,  \cite{LS} section 8.6,   for  a  continuous 
potential $\psi(x)$  the Dirac operator is always  in the limit point  case. It means that for $\l$ with $\Im \l\neq 0$   there exists one  solution from $\LLL^2[y,+\infty)$ and another solution from  $\LLL^2(-\infty,y]$. Evidently the Weyl solutions are determined up to a multiplicative constant.

Pick some  $\alpha \in [0,\pi)$. Consider the fundamental system of  
solutions $\pp_{\alpha}(x,y,\l)$ and
$\uu_{\alpha}(x,y,\l)$ of \ref{spro}  normalized by
\beq\label{norm}
\pp_{\alpha}(x,y,\l)|_{x=y}=\[\begin{matrix} 
ie^{i\alpha}\\
-ie^{-i\alpha}
\end{matrix}\], \qquad\qquad\qquad 
\uu_{\alpha}(x,y,\l)|_{x=y}=\[\begin{matrix}
e^{i\alpha}\\
e^{-i\alpha}
\end{matrix} \]. 
\eeq
The   Weyl solution $\e(x,y,\l)$ from $\LLL^2[y,+\infty)$ or $\LLL^2(-\infty,y]$ is proportional to a linear combination of 
$\pp_\alpha$ and $\uu_\alpha$: 
$$\e\sim a \pp_{\alpha} +b \uu_{\alpha},
$$
but  the {\it Weyl function} $\P_{\alpha}^\pm(y,\l)=a/b$ 
is defined uniquely. Evidently, the solution $\P_{\alpha}^\pm\pp_{\alpha}+ \uu_{\a}
$ 
belongs to  $\LLL^2[y,+\infty)/\LLL^2(-\infty,y]$ for $\l$ with $\Im \l \neq 0$. 

\begin{exa} {\bf The trivial potential $\psi(x)\equiv 0$.} \end{exa}
$$\pp_{\alpha}(x,y,\l) = \left[\begin{matrix}
i e^{i\alpha-{i\l\over 2}(x-y)}\\
-ie^{-i\alpha +{i\l\over 2}(x-y)}
\end{matrix}  \right]
,  \qquad \qquad
\uu_{\alpha}(x,y,\l) = \left[\begin{matrix}
e^{i\alpha -{i\l\over 2}(x-y)}\\
e^{-i\alpha+ {i\l\over 2}(x-y)}
\end{matrix}  \right].
$$
The Weyl function $\P_{\alpha}^+(y,\l)=i$ if $\Im \l >0$ and $-i$ for $\Im \l <0$. 
The Weyl function $\P_{\alpha}^-(y,\l)=-i$ if $\Im \l >0$ and $i$ for $\Im \l <0$.
The only potential which has such Weyl functions 
vanishes identically. 

\begin{lem}\label{rotl}
Any two   functions $\P_{\alpha}^\pm(y,\l)$ and $\P_{\beta}^\pm(y,\l)$ are
related by the equation 
\beq\label{rot}
\P_{\alpha}^\pm={\P_{\beta}^\pm\cos(\alpha-\beta)-\sin(\alpha-\beta)\over
\P_{\beta}^\pm\sin(\alpha-\beta)+ \cos(\alpha-\beta)}.  
\eeq
\end{lem}

{\it Proof.} First we will obtain the expression for the function $\P_{\alpha}^\pm(y,\l)$
in terms of  the  square integrable solution $\e(x,y,\l)$. 
This solution is proportional to $\P_{\alpha}^\pm\pp_{\alpha}+ \uu_{\a}$ with some constant $c$. 
Thus $\P_{\alpha}^\pm \pp_{\alpha}+ \uu_{\alpha} = c\e$ and  normalization conditions \ref{norm} imply the system
\bey
\P_{\alpha}^\pm(y,\l) i e^{i\alpha}+ e^{i\alpha} &=&c e_1(y,y,\l),\\
 -\P_{\alpha}^\pm(y,\l) i e^{-i\alpha}+e^{-i\alpha}&=&c e_2(y,y,\l).
\eey
Solving for $\P_{\alpha}^\pm$,  
\beq\label{imp}
\P_{\alpha}^\pm={e_1e^{-i\alpha}-e_2e^{i\alpha}\over e_1ie^{-i\alpha}
+e_{2}i e^{i\alpha}}.  
\eeq
This identity can be written as
$$
\P_{\alpha}^\pm=\frac{(e_1 e^{-i\b}-e_2e^{i\b})\cos(\a-\b) -(e_1ie^{-i\b}+e_2 
ie^{i\b})\sin(\a-\b)}{ (e_1 i e^{-i\b}+e_2ie^{i\b})\cos(\a-\b) +
(e_1e^{-i\b}-e_2 e^{i\b})\sin(\a-\b)}.
$$
Dividing the numerator and denominator by $e_1 i e^{-i\b}+e_2ie^{i\b}$ 
we obtain \ref{rot}. 
\qed

\noi
Therefore, if $\P_{\alpha}^\pm$ is known for some value of the parameter 
$\alpha$, then it  is known for all other values of  $\alpha$.

If $\f(x,\l)$ is a solution of the auxiliary problem \ref{spro}  
corresponding to $\l$, then $\hat \f=\sigma_1 \overline{\f}$ is a solution   corresponding to $\overline{\l}$. Formula \ref{imp} implies, for $\l$ with 
$\Im \l \neq 0$,
$$
\P_\alpha^\pm(\lb)={e_1(\lb)e^{-i\alpha}-e_2(\lb)e^{i\alpha}\over e_1(\lb)ie^{-i\alpha}
+e_{2}(\lb)i e^{i\alpha}}={\overline{e}_2(\l)e^{-i\alpha}-\overline{e}_1(\l)e^{i\alpha}\over \overline{e}_2(\l)ie^{-i\alpha}
+\overline{e}_{1}(\l)i e^{i\alpha}}=\overline{\P}_\alpha^\pm(\l).
$$
The function $\P_\alpha^\pm(y,\l)$ for fixed $y$ maps the upper/lower half-plane into the upper/lower half-plane.

For a nonzero potential the Weyl  function $\P_{\alpha}^+(y,\l)$ takes valus in the upper hulf--plane 
  when  $\l$ is in  the  upper half--plane.  
It is represented   by the integral, \cite{KK},
$$
\P_{\alpha}^+(y,\l)= b\l +a + \int\limits^{+\infty}_{-\infty}
 \[{1\over t-\l}- {t\over t^2+ 1}\] 
d\sigma_{\alpha}(t), 
$$ 
with
$$
b \geq 0,\qquad a \in \RB^1,\qquad \int\limits^{+\infty}_{-\infty}
{d \sigma_{\alpha}(t)\over t^2 +1} < \infty. 
$$
The  spectral  measure $d\sigma_{\alpha}$ corresponds to a  self--adjoint extension 
of the operator  $\Dg$    on $\LLL^2[y,+\infty)$  specified by the  boundary  condition
$$
f_1(y) e^{-i\alpha}=f_2(y) e^{i\alpha}.
$$
The function $\P_{\alpha}^-(y,\l)$ has a similar representation.

For each point of the complex plane  with nonzero imaginary part there are two Weyl solutions. To make the Weyl solution a single valued 
function of a point we introduce $\G=\G_{+} \cup \G_{-}$, a two sheeted covering of the complex plane (Figure 1). Each sheet $\G_{+}$ or $\G_{-}$ is a 
copy of the complex plane without the real line. 
Each point  of the cover $\G$ is a pair $Q=(\l,\pm)$ where $\l$ is a point of the complex plane and the sign $\pm$ specifies the sheet. 
We denote by $P_+$ and $P_-$ the  infinity corresponding to the   sheet $\G_{+}$ or $\G_{-}$ (Figure 2). 
\begin{figure}[htb]
\includegraphics[width=0.60\textwidth]{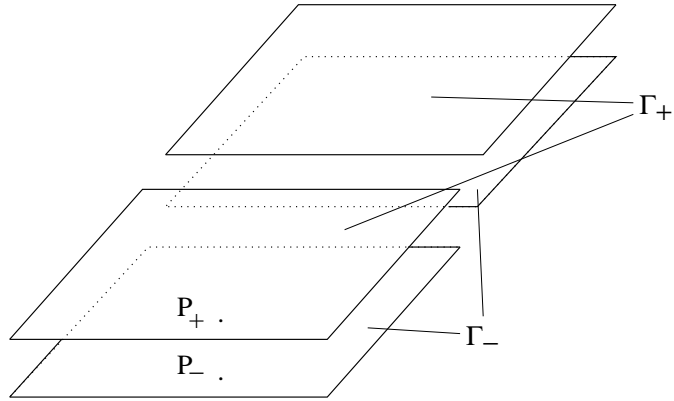}
\caption{ Two sheets of the spectral cover.}
\end{figure}

Let us introduce two components of the spectral cover
$$
\G_R=\{Q\in \G_+,\; \Im \l(Q) >0\} \cup \{Q\in \G_-,\; \Im \l(Q) <0\}
$$
and 
$$
\G_L=\{Q\in \G_+,\; \Im \l(Q) <0\} \cup \{Q\in \G_-,\; \Im \l(Q) > 0\}.
$$
Evidently $\G=\G_R\cup \G_L$. 
On $\G$ we define an involution $\ea$ by the rule 
$$
\ea:\quad (\l,\pm)\longrightarrow (\lb,\mp).
$$
The involution permutes infinities $\ea: P_+\longrightarrow P_-$. Evidently $\G_R$ or $\G_L$ are invariant under the action of $\ea$. 

We define the Weyl solution $\e(x,y,Q),\, x,y \in \RB^1,\; Q\in \G$, which is the vector function 
$$
\e(x,y,Q)=\[\begin{matrix}
e_1(x,y,Q)\\
e_2(x,y,Q)
\end{matrix} \],
$$
as a solution of \ref{spro} with $\l=\l(Q)$ and the following properties
\begin{itemize}
\item  $\e(x,y,Q)$ belongs to $\LLL^2[y,+\infty)$  if  $Q \in \G_R$;
\item  $\e(x,y,Q)$ belongs to $\LLL^2(-\infty,y]$  if  $Q \in \G_L$;
\end{itemize} 
normalized by the condition 
\beq\label{normn}
e_1(x,y,Q)|_{x=y}=1.
\eeq

The standard transition  matrix  $M(x,y,\l)=\(M^{(1)},M^{(2)}\)$ is  
$2\times2$ matrix solution of the auxiliary linear problem  \ref{spro}   
which satisfies  the boundary condition $M(x,y,\l)|_{x=y}=I$. The symmetry of the matrix $V(x,\l)$
$$
\sigma_1 \overline{V}(x,\l)\sigma_1= V(x,\bar{\l})
$$
produces  the same relation for the transition  matrix 
\beq\label{rreal}
\sigma_1 \overline{M}(x,y,\l)\sigma_1= M(x,y,\bar{\l}).
\eeq
This implies  for the columns
\bay\label{realc}
M^{(1)}(x,y,\lb)=\sigma_1\overline{M}^{(2)}(x,y,\l),\qquad\qquad
M^{(2)}(x,y,\lb)=\sigma_1\overline{M}^{(1)}(x,y,\l).
\ey
Evidently the Weyl solution $\e(x,y,Q)$ is a linear combination of the columns of the transition matrix. The normalization condition \ref{normn}
implies 
\beq\label{expl}
\e(x,y,Q)=M^{(1)}(x,y,\l)+\P(y,Q)M^{(2)}(x,y,\l),
\eeq
where the function $\P(y,Q)$ is called the  Weyl function.

Formulas \ref{realc},     \ref{expl}  and the uniqueness of the Weyl function   imply 
\beq\label{reln}
\P(y,\ea Q)=\frac{1}{\overline{\P(y,Q)}}.
\eeq

\begin{exa}\label{exatr} {\bf The trivial potential $\psi(x)\equiv 0$.} \end{exa}
\noi
For the columns of the transition matrix we have 
$$
M^{(1)}(x,y,\l)= e^{-i \lt(x-y)} 
\[\begin{array}{ccccc}  1\\ 0\end{array}\], \qquad\qquad 
M^{(2)}(x,y,\l)= e^{+i \lt(x-y)} \[\begin{array}{ccccc}
0\\
1\end{array} \].
$$
We put $\P(y,Q)=\infty$ when $Q\in \G_+$ and $\P(y,Q)=0$ when $Q\in \G_-$.

For  $\psi(x)\neq 0$ the function $\P(x,Q)$  has a pole at $P_+$ and a zero at $P_-$. 

Arguing as in the proof of Lemma \ref{rotl},  for $Q\in \G_R$ we have 
\beq\label{rel}
\P(y,Q)={i+\P_{\a}^+(y,\l)\over i-\P_{\a}^+(y,\l)}e^{-i2\alpha}, \qquad\qquad\qquad \l=\l(Q).
\eeq 
Note that the  transformation
$$
z\longrightarrow z'={i+z\over i-z}e^{-i2\alpha}
$$
establishes a 1:1 correspondence between the upper/lower  half--plane and the
exterior/interior of the unit circle.
This  implies, in particular, that the function $\P(y,Q)$ for  $Q\in \G_R$ with $\Im Q >0$ or $\Im Q <0$
takes values in the exterior or interior of the unit circle.   The shift $\alpha\longrightarrow \beta$
described on the $z$--plane by  formula \ref{rot}  corresponds to 
rotation via the angle $2(\alpha-\beta)$ on the $z'$--plane.

For $Q\in \G_L$ we  also have 
\beq\label{rell}
\P(y,Q)={i+\P_{\a}^-(y,\l)\over i-\P_{\a}^-(y,\l)}e^{-i2\alpha}, \qquad\qquad\qquad \l=\l(Q).
\eeq 
The function $\P(\l,Q)$ for  $Q\in \G_L$ with $\Im Q < 0$ or $\Im Q >0$
takes values in the exterior or interior of the unit circle.   

We associated to a potential $\psi(x)$ defined on the entire line  a pair $(\G, \P)$. Thus we defined a map 
\beq\label{dst}
\MM \qquad\longrightarrow 
\qquad (\G, \P),
\eeq
which we call {\it the direct spectral transform}. This map is injective. 
If the function $\P(y,Q)$ is known for some fixed value of $y$,  formulas \ref{rel}-\ref{rell} allow one  to reconstruct $\P_{\a}^-(y,\l)$ and  $\P_{\a}^+(y,\l)$ (or equivalently the spectral measure).  
By the theorem of Marchenko, \cite{MA1}, a potential is determined  on the corresponding half line  by its spectral measure. 

Due to its injective character the direct spectral transform can be inverted
\beq\label{ist}
\qquad (\G, \P)\qquad\longrightarrow \qquad \MM.
\eeq
This map we call {\it the inverse spectral transform}. 
An effective procedure for  the inverse spectral transform was constructed by Gelfand and Levitan, \cite{GL}.

\subsection{\bf The formal  series for the  Weyl functions.}\label{asym}

For the function $\P(y,Q),\, Q\in \G_+,\; \Im Q >0,$  and $z < y$ we have  
$$
\P(y,Q)=\frac{e_2(y,y,Q)}{e_1(y,y,Q)}=\frac{e_2(y,z,Q)}{e_1(y,z,Q)}.
$$
This together with \ref{spro} implies the Ricatti type equation, $\l=\l(Q)$: 
\beq\label{Ricm}
{d\P(y,Q)\over dy}= i\l \P(y,Q) + \psi -\psib \P^2(y,Q). 
\eeq
In fact, the function $\P(y,Q)$ satisfies the same equation in the variable $y$ for any $Q$. 

For a generic potential the Weyl function $\P(y,Q)$ has a pole at infinity for $Q\in \G_+$ (compare Example \ref{exatr}). The formal  expansion has the form 
$$
\P(y,Q)= {a_{-1}(y)}\l +a_0(y)+\frac{a_1(y)}{\l}+\hdots,   \qquad\qquad \qquad Q\in (P_+); 
$$
with
$$
a_{-1}(y)=\frac{1}{-i\psib(y)},  \qquad\qquad    
a_0(y)=\frac{\psib'(y)}{\psib^2(y)},     \qquad   \qquad etc.
$$
The function $\P(y,Q)$    has a zero at infinity for $Q\in \G_-$. It has the  power series expansion  
$$
\P(y,Q)= \frac{b_1(y)}{\l} +\frac{b_2(y)}{\l^2}+\frac{b_3(y)}{\l^3}+\hdots, \qquad\qquad \qquad Q\in (P_-);
$$
where 
$$
b_1(y)=i\psi(y),  \qquad\qquad   
b_2(y)= \psi'(y),  \qquad   \qquad etc.
$$
The coefficients of the two expansions are obtained by substituting the series  into  the differential equation \ref{Ricm} and then by matching terms with the same power of $\l$. The  expansions are connected in accordance with  formula \ref{reln}.  

V. A. Marchenko, \cite{MA2}, studied the question of when the formal series have an asymptotic character. If the potential has some degree of differentiability then the formal series have an asymptotic character.   According to \cite{MA2} a converse  statement is also true. If the formal series converge for  large values of  $\l$, then the corresponding potential is an infinitely differentiable function. In fact, for our purposes it is enough to have only the  first term of the  asymptotic  expansion.

The Weyl functions $\P_{\a}^+(y,\l)$ and $\P^-_{\a}(y,\l)$ for fixed $\l$ and $\a$ are 
the functions of $y$ and the potential. 
Formula  \ref{imp}  and the spectral problem  \ref{spro} imply the Ricatti--type
equation  for  $\P_{\a}^\pm(y,\l)$:
$$
{d\P_{\a}^\pm(y,\l)\over dy}=-{\l\over2} ({\P_{\a}^\pm}^2(y+,\l)+1 ) - \P_{\a}^\pm(y,\l)
(\psib_\alpha +\psi_\alpha) + {i\over 2} ({\P_{a}^\pm}^2(y,\l) + 1)
(\psib_\alpha -\psi_\alpha),
$$
where $\psi_\alpha=\psi e^{i2\alpha}$. In general, the function $\P(y,Q)$ is simpler to work with than $\P_\a^\pm(y,\l)$. 

\section{The Poisson bracket.}

\subsection{\bf The Atiyah--Hitchin  bracket for the Weyl function.} 
The main result of this section is Theorem \ref{ahb}. This theorem  describes the image of the Poisson bracket \ref{nlspb}  under the direct spectral transform \ref{dst}.

\begin{thm}\label{ahb} Let $y$ be some fixed point of the line and $\P(Q)=\P(y,Q)$ and  $\P(P)=\P(y,P)$.

i. If $Q$ and $P$ belong to  $\G_R$, then
$$
\{\P(Q), \P(P)\}=
2\times {(\P(Q)-\P(P))^2\over \l(Q) - \l(P)}.
$$ 

ii. If $Q$ and $P$ belong to  $\G_L$, then
$$
\{\P(Q), \P(P)\}=
-2\times {(\P(Q)-\P(P))^2\over \l(Q) - \l(P)}.
$$ 
iii. If  $Q$  belongs to  $\G_R$, $P$ belongs to $\G_L$, then 
$$
\{\P(Q), \P(P)\}=0.
$$
\end{thm}

\begin{rem}
In the formulation of Theorem  we have to consider various parts of  the spectral cover  separately because it is not simply--connected. 
\end{rem}

Formulas \ref{rel} and \ref{rell}  and the invariance of the AH bracket under linear fractional transformations (Lemma \ref{invan})  
produce the bracket  for  $\P_\alpha^\pm(y,\l)$.  We precede the proof  with two auxiliary results.

\begin{lem} \label{iden}
Let the  vectors $\f^\heartsuit(x,\l),\; \f^\spadesuit(x,\l)$ satisfy 
$$
{\f^\heartsuit}'(x,\l)=V(x,\l)\f^\heartsuit(x,\l), \qquad\qquad\qquad {\f^\spadesuit}'(x,\l)=V(x,\l)\f^\spadesuit(x,\l),
$$
and the  vectors $\gb^\heartsuit(x,\m),\; \gb^\spadesuit(x,\m)$ satisfy
$$
{\gb^\heartsuit}'(x,\m)=V(x,\m)\gb^\heartsuit(x,\m), \qquad\qquad\qquad {\gb^\spadesuit}'(x,\m)=V(x,\m)\gb^\spadesuit(x,\m).
$$ 
The following identity holds:
\bey\label{idt}
f_1^\heartsuit f_1^\spadesuit  g^\spadesuit_2 g_2^\heartsuit- f_2^\spadesuit f_2^\heartsuit g_1^\heartsuit g_1^\spadesuit= {1\over i(\m-\l)}
\times {d\over dx}
\[ \( {\f^\heartsuit}^TJ \gb^\heartsuit\) \( {\f^\spadesuit}^T J \gb^\spadesuit\)\].
\eey
\end{lem} 
\noi
{\it Proof.} The identity can be verified by differentiation. \qed

\begin{lem}\label{grad}
If $Q\in \G_R$, then the following identities hold:
\bay
{\d \P (z,Q)\over \d \psi(y)} &=&  
-\left[e_1(y,z,Q)\right]^2,\qquad\qquad  \qquad y \geq z;\label{first}\\
{\d \P(z,Q) \over \d \psib(y)} &=&  \quad
\left[e_2(y,z,Q)\right]^2,\qquad\qquad  \qquad y \geq z. \label{sec}
\ey
The derivatives vanish for $y < z$.

\noi 
If $Q\in \G_L$, then the following identities hold:
\bay
{\d \P (z,Q)\over \d \psi(y)} &=& \quad 
\[e_1(y,z,Q)\]^2,\qquad\qquad   y \leq z;\label{nthi}\\
{\d \P(z,Q) \over \d \psib(y)} &=&  -
\[e_2(y,z,Q)\]^2,\qquad\qquad  y\leq z.\label{nfou}
\ey
The derivatives vanish for $y>z$. 
\end{lem}

The lemma was proved first in \cite{V6}. Here 
we present a simplified  proof  of    formulas \ref{first} and \ref{sec}. 
The  proof of  other formulas \ref{nthi} and \ref{nfou} is  the same.
We split the proof into small steps.

\noi{\it Proof.} Without loss of generality we assume that $z=0$. 

\noi
{\it Step 1.} This step prepares gradients. 
Let $M^{\bullet}=\d M$ be a variation of $M(x,0)$ in response to the variation 
of $\psi(y)$ and $\psib(y),\; 0\leq y\leq x$. Then 
$M^{{\bullet}'}=VM^{\bullet} + V^{\bullet}M$. The solution of this 
nonhomogenious equation is  
$$
M^{\bullet}(x)=M(x)\int\limits_{0}^{x} M^{-1}(\xi)V^{\bullet}(\xi) 
M(\xi)\,d\xi.
$$
Therefore, 
$$
{\d M(x,0)\over \d \psi(y)}=M(x,y)\left(\begin{matrix} 0& 0\\
                                                            1& 0 \end{matrix} \right) M(y,0),
$$
$$
{\d M(x,0)\over \d \psib(y)}=M(x,y)\left(\begin{matrix} 0& 1\\
                                                             0& 0 \end{matrix}
\right) M(y,0). 
$$
 
\noi
{\it Step 2.} 
The purpose of this  step is to prove the formula
\beq\label{nidenim}
{\d \P(0,Q)\over \d \psi(y)}=
\P(0,Q)\left[{m_{11}(y,0)\over
A}+{m_{12}(y,0)\over B}\right],
\eeq
where
\bey
A&=& \;\;\; m_{21}(y,0)-\P(y, Q)m_{11}(y,0),\\
B&=& -m_{22}(y,0)+ \P(y, Q) m_{12} (y,0). 
\eey
 
Consider the auxiliary spectral  problem  \ref{spro}  on the  finite 
interval $\[a,b\]$.  The solution 
$$
\f(x,\lambda)= M^{(1)}(x,a,\lambda)+  \P(b,a,Q) M^{(2)}(x,a,\lambda),
$$
with some $\P(b,a,Q)$, satisfies the boundary condition   
$f_1(b,\lambda)=f_2(b,\lambda)$ if
$$
\P(b,a,Q) ={m_{11} -m_{21}\over m_{22}-m_{12}}(b,a,\lambda).
$$

The limit
$$
\P(0,Q)=\lim\limits_{b\rightarrow +\infty} \P(b,0,Q)
$$
exists because the spectral problem is in the limit--point case.  Therefore,\footnote{$\nabla=\d/\d \psi(y)$ or $\d/\d \psib(y)$.}  
$$
\nabla \P(0,Q)=\lim\limits_{b\rightarrow +\infty} \nabla \P(b,0,Q). 
$$
To compute the derivative
\bay
\nabla \P(b,0,\l)=&&{m_{11} -m_{21}\over m_{22}-m_{12}}
                      {\nabla m_{11} -\nabla m_{21}\over m_{11}-m_{21}}
(b,0,\lambda)\nonumber\\
 & -& {m_{11} -m_{21}\over m_{22}-m_{12}}
    {\nabla m_{22} -\nabla m_{12}\over m_{22}-m_{12}}(b,0,\lambda).\label{der} 
\ey
we use the formulas of Step 1:
\bey
{\d \P(b,0,Q)\over \d \psi(y)} =& \;&\P(b,0,Q)
                         {[m_{12} -m_{22}](b,y)m_{11}(y,0)\over [m_{11}-m_{21}](b,0)} \\
          & -& \P(b,0,Q) {[m_{22} -m_{12}](b,y)m_{12}(y,0)\over 
[m_{22}-m_{12}](b,0)} \\
                     =& \;&\P(b,0,Q)[m_{12}-m_{22}](b,y)\\
             & \times&  \left[{m_{11}(y,0)\over [m_{11}-m_{21}](b,0)}
              + {m_{12}(y,0)\over [m_{22}-m_{12}](b,0)} \right].
\eey
Using the identity $M(b,0)=M(b,y)M(y,0)$ and simple algebra, one finds
\bey
\P (b,0,Q)
 &\times & \left[  {m_{11}(y,0)\over m_{21}(y,0) - \P(b,y,Q)m_{11}(y,0) }
+{m_{12}(y,0)\over
-m_{22}(y,0) +\P(b,y,Q)m_{12}(y,0)} \right]. 
\eey
Now pass to the limit when $b\rightarrow \infty$.

\noi
{\it Step 3.} 
The purpose of this  step is to prove the formula
\beq\label{identt}
{\d \P(0,Q)\over \d \psib(y)}=
- \P(0,Q)\P(y,Q)\left[{m_{21}(y,0)\over
A}+{m_{22}(y,0)\over B}\right].
\eeq
To compute the derivative $\nabla \P(b,0,\l)$  given by \ref{der}
we use the formulas of Step 1
\bey
{\d \P(b,0,Q)\over \d \psib(y)} =& \;&\P(b,0,Q)
                         {[m_{11} -m_{21}](b,y)m_{21}(y,0)\over [m_{11}-m_{21}](b,0)} \\
          & -& \P(b,0,Q) {[m_{21} -m_{11}](b,y)m_{22}(y,0)\over 
[m_{22}-m_{12}](b,0)} \\
                     =& \;&\P(b,0,Q)[m_{11}-m_{21}](b,y)\\
             & \times&  \left[{m_{21}(y,0)\over [m_{11}-m_{21}](b,0)}
              + {m_{22}(y,0)\over [m_{22}-m_{12}](b,0)} \right].
\eey
Using the identity $M(b,0)=M(b,y)M(y,0)$,  and simple algebra, one finds
\bey
&+&\P (b,0,\l) \P(b,y,\l)\\
 &\times & \left[  {m_{21}(y,0)\over -m_{21}(y,0) + \P(b,y,\l)m_{11}(y,0) }
+{m_{22}(y,0)\over
m_{22}(y,0) -\P(b,y,\l)m_{12}(y,0)} \right]. 
\eey
Now pass to the limit when $b\rightarrow \infty$.

\noi
{\it Step 4.} 
Consider $\e(x,y,Q)$  proportional to  $\e(x,0,Q)$. Then 
\beq\label{del}
\P(y,Q)={e_2(y,y,Q)\over e_1(y,y,Q)}={e_2(y,0,Q)\over e_1(y,0,Q)}.  
\eeq
Therefore, 
$$
\P(y,Q)[m_{11}(y,0)+\P(0,Q)m_{12}(y,0)]=
m_{21}(y,0)+\P(0,Q)m_{22}(y,0).
$$
After simple algebra,
\beq\label{ident}
{\P(0, Q)\over A}={1\over B}. 
\eeq

Therefore, using \ref{nidenim}, \ref{del}, and \ref{ident}, we obtain 
\bey
{\d \P(0,Q)\over \d \psi(y)}& =&\P(0,Q)
\left[{ m_{11}(y,0)+ \P(0,Q)m_{12}(y,0)\over A}\]\\
&=&\P(0,Q){e_1(y,0,\lambda)\over m_{21}(y,0)-\frac{e_2(y,0,Q)}{e_1(y,0,Q)} m_{11}(y,0)}\\
&=&\P(0,Q) { \left[e_1(y,0,Q)\right]^2\over m_{21}(y,0)e_1(y,0,Q)-m_{11}(y,0)e_2(y,0,Q)}. 
\eey
The denominator does not depend on $y$ and  can be computed for  $y=0$, 
where it is equal to $-\P(0,Q)$. Formula \ref{first} is proved.

\noi
{\it Step 5.} 
Using \ref{identt}, \ref{del} and \ref{ident}  we obtain 
\bey
{\d \P(0,Q)\over \d \psi(y)}& =&-\P(0,Q)\P(y,Q)
\left[{ m_{21}(y,0)+\P(0,Q) m_{22}(y,0)\over A}\]\\
&=&-\P(0,Q)\P(y,Q){e_2(y,0,Q)\over m_{21}(y,0) - \P(y,Q)m_{11}(y,0)}\\
&=&-\P(0,Q){ \left[e_2(y,0,Q)\right]^2\over m_{21}(y,0)e_1(y,0,Q)-m_{11}(y,0)e_2(y,0,Q)}. 
\eey
The denominator does not depend on $y$ and  can be computed for  $y=0$, 
where it is equal to $-\P(0,Q)$. Formula \ref{sec} is proved. 
 \qed

Now we are ready to prove the theorem.

\noi
{\it Proof.}  We will prove the first formula. All others can be treated the same way. By  Lemma \ref{grad}, 
\bey
&\phantom{ooo}&\{\P(Q),\P(P)\}=\\
& =&  2i\int\limits_{y}^{+\infty}
{\d \P(y,Q)\over \d \psib(\xi)}{\d \P(y,P)\over \d \psi(\xi)}-
{\d \P(y,Q)\over \d \psi(\xi)}{\d \P(y,P)\over \d \psib(\xi)}\, d\xi\\
                 &=&   2i\int\limits_{y }^{+\infty} -e_2^2(\xi,y,Q)
e_1^2(\xi,y,P) + e_1^2(\xi,y,Q)e_2^2(\xi,y,P) \, d\xi.
\eey
Using the identity of Lemma \ref{iden}, 
\bey
 \therefore&=&   {2i\over i(\l(P)-\l(Q))} \[\e(\xi,y,Q)^TJ \e(\xi,y,P)\]^2|_{y }^{+\infty}\\
                 &=& 2\times {(\P(Q)-\P(P))^2\over \l(Q) - \l(P)}.
\eey
The  formula  is proved. \qed

\subsection{\bf Computation of the Poisson bracket for field variables.}

The inverse spectral transform \ref{ist}  maps the AH bracket on  Weyl functions given by Theorem \ref{ahb} to the  phase space.  
The main result of this section is the following 
\begin{thm}\label{ppbb}
The  AH bracket for the field variables $\psi(x)$ and $\psib(x)$ is given by the formulas\footnote{ The identities are understood in the sence of generalized functions: $u(x)=v(x)$ if for any $f(x)\in C^{\infty}_0$ we have $\int u(x) f(x) dx= \int v(x) f(x) dx$. }:
\bay
\{\psi(z),\psi(y)\}&=&0,\label{pp}\\
\{\psib(z),\psib(y)\}&=&0,\label{pop}\\
\{\psib(z),\psi(y)\}&=&2i \d(z-y).\label{thi}
\ey
These identities are an equivalent form of the Poisson bracket \ref{nlspb}. 
\end{thm}

The inverse spectral transform \ref{ist} is very implicit.  However, if we know $\P(y,Q)$ for all values of the variable y, then the series presented in  \ref{asym} imply 
\beq\label{psib} 
\lim_{Q \rightarrow P_+} \frac{\l(Q)}{\P(y,Q)}=-i\psib(y)
\eeq
and 
\beq\label{psi}
\lim_{Q \rightarrow P_-} \l(Q)\P(y,Q)=i\psi(y).
\eeq
The limits are complex conjugate of each other due to \ref{reln} and 
\bey
\lim_{Q\rightarrow P_-} \l(Q) \P(x,Q)&=& \lim_{Q\rightarrow P_-} \l(\ea \ea Q) \P(x,\ea \ea Q)=
\lim_{Q\rightarrow P_-} \overline{\frac{\l( \ea Q)} {\P(x, \ea Q)}}\\
&=&\lim_{Q\rightarrow P_+} \overline{\frac{\l(  Q)} {\P(x,  Q)}}.
\eey
To prove  theorem \ref{ppbb} using these formulas   one needs  to compute the bracket 
$
\{\P(x,Q),\P(y,P)\}
$
when $x\neq y$. Fortunately one needs this formula only  asymptotically when one of the points $Q$ or $P$  tends to  infinity. 

\begin{lem}\label{rrt}  Suppose $Q \rightarrow  P_{\pm}$ from the imaginary direction. Then 
\beq\label{asahd}
\{\P(y,Q),\P(x,P)\}\sim  e^{-i\l(Q)(x-y)}\{\P(x,Q),\P(x,P)\}. 
\eeq
\end{lem}
 
\noi
{\it Proof.} The identity   
$$
\e(y,z,Q)=M(y,x,\l) \e(x,z,Q),\qquad\qquad\qquad \l=\l(Q), 
$$ 
implies  
$$
\P(y,Q)=\frac{m_{22}(y,x,\l)\P(x,Q)+ m_{21}(y,x,\l)}{m_{12}(y,x,\l)\P(x,Q)+ m_{11}(y,x,\l)}.
$$ 
If  $\l=i\tau,\; \tau\rightarrow \pm \infty$, then   
$$
M(y,x,\l)\sim e^{-i\l/2(y-x)\sigma_3},
$$
{\it i.e.}, the transition matrix near infinity behaves like the solution of the free ($Y_0=0$) equation.
To see this write the integral equation for the transition  matrix
$$
M(y,x,\l)=e^{-i\l/2(y-x)\sigma_3}   +\int_{x}^{y}d\xi e^{-i\l/2(y-\xi)\sigma_3} Y_0(\xi) M(\xi,x,\l).
$$
In symbolic form $M=R+AM$, where $R$ is the  solution of the free equation, and $A$ is the integral operator. 
Now expand the solution into the Neumann series 
$$
M=R+AR+A^2R+\hdots.
$$
and take the first term of the expansion. This produces the stated asymptotics.  
\qed

\begin{rem}
We cut the tail of the Neumann series. An  estimate for the tail for infinitely smooth potentials is given in \cite{V44} and for 
general square integrable potentials in \cite{MCV}. 
\end{rem}
 
The next lemma establishes that the Poisson tensor is real. 
\begin{lem}\label{pbreal} The Poisson brackets for the field variables $\psi(x)$ and $\psib(x)$ are real 
\bey
\overline{\{\psi(y),\psi(z)\}}&=&\{\psib(y),\psib(z)\},\label{firre}\\
\overline{\{\psib(y),\psi(z)\}}&=&\{\psi(y),\psib(z)\}.\label{secre}
\eey
\end{lem}

\noi
{\it Proof.}  We will prove the first identity. The second identity can be proved along the same lines.

For  $Q,P \in \G_R,$ using \ref{psi}, we have  
\bey
\overline{\{\psi(y),\psi(z)\}}&=&-\overline{\{i\psi(y),i\psi(z)\}}\\
&=&-\lim_{Q,P\rightarrow P_-} \overline{\{\l(Q)\P(y,Q),\l(P)\P(z,P)\}}\\
&=& -\lim_{Q,P\rightarrow P_-}\lb(Q)\lb(P) \overline{\{\P(y,Q),\P(z,P)\}}.
\eey
We assume that  the limit is taken along the imaginary direction.
Using Lemma \ref{rrt},
$$
\therefore\quad=  -\lim_{Q,P\rightarrow P_-}\lb(Q)\lb(P) e^{i\bar{\l}(Q)(z-y)}\overline{\{\P(z,Q),\P(z,P)\}}
$$
By Theorem \ref{ahb}, 
$$
\therefore\quad= -\lim_{Q,P\rightarrow P_-}\lb(Q)\lb(P) e^{i\bar{\l}(Q)(z-y)}\times  2 {(\overline{\P}( z,Q)-\overline{\P}( z,P))^2\over \lb(Q) - \lb(P)}.
$$
Now we use $Q=\ea \ea Q,\, P=\ea \ea P$ and  invariance of $\G_R$ under  the action of $\ea$. From formula \ref{reln}, 
after simple algebra
\bey
\therefore&=&-\lim_{Q,P\rightarrow P_-}\frac{\l(\ea Q)\l(\ea P)e^{i{ \l}(\ea Q)(z-y)}}{\P^2( z,\ea Q)\P^2( z,\ea P)}\times 2 {(\P( z,\ea Q)-\P( z,\ea P))^2\over \l(\ea Q) - \l(\ea P)}\\
&=&-\lim_{Q,P\rightarrow P_+}\frac{\l( Q)\l( P)e^{i{ \l}( Q)(z-y)}}{\P^2( z, Q)\P^2( z, P)}\times 2 {(\P( z, Q)-\P( z, P))^2\over \l( Q) - \l( P)}\\
&=& -\lim_{Q,P\rightarrow P_+} \l( Q)\l( P)e^{i{ \l}( Q)(z-y)} \{\frac{1}{\P( z, Q)},\frac{1}{\P( z, P)}\} \\
&=& -\lim_{Q,P\rightarrow P_+} \l( Q)\l( P) \{\frac{1}{\P( y, Q)},\frac{1}{\P( z, P)}\}=-\{-i\psib(y),-i\psib(z)\} .
\eey
The last line follows from \ref{psib}.  
\qed

Now we are ready to prove the main result.

\noi
{\it Proof of Theorem \ref{ppbb}.}   Due to Lemma \ref{pbreal},  identities \ref{pp} and \ref{pop} are equivalent. 
We compute  the bracket \ref{pp}. Let $Q,P\in \G_R$ and  $Q\rightarrow P_-,\; P\rightarrow P_-$ along the imaginary direction. Using  formula \ref{psi} and Lemma \ref{rrt} for  $f(x)\in C^{\infty}_0$ and $y \leq z$ we have: 
\bey
&\phantom{000}&\int\limits_{-\infty}^{z} dy  f(y) \{\psi(z),\psi(y)\}=-\int\limits_{-\infty}^{z} dy f(y) \{i\psi(z),i \psi(y)\}\\
&=&-\lim_{Q,P\rightarrow P_-} \int\limits_{-\infty}^{z} dy f(y) \{\l(Q)\P(z,Q),\l(P)\P(y,P)\}\\
&=& -\lim_{Q,P\rightarrow P_-}  \l(Q)\l(P)\int\limits_{-\infty}^{z} dy f(y) \{\P(z,Q),\P(y,P)\}\\
&=& -\lim_{Q,P\rightarrow P_-}  \l(Q)\l(P) \{\P(z,Q),\P(z,P)\}\int\limits_{-\infty}^{z} dy f(y)e^{-i\l(P)(z-y)}.\\
\eey
Let $Q,P$ be such that $\l(Q)=-i\tau,\;\l(P)=-2i\tau $.
Since $\P$ has a zero at $P_-$ we have, using Theorem \ref{ahb} when $\tau \rightarrow +\infty$, 
\bey
\lim_{Q,P\rightarrow P_-}  \l(Q)\l(P) \{\P(z,Q),\P(z,P)\}\sim \l(Q)\l(P) {(\P(z,Q)-\P(z ,P))^2\over \l(Q) - \l(P)}=O(\tau^{-1}). 
\eey
For the integral   we have 
\bey
\int\limits_{-\infty}^{z} dy f(y)e^{-i\l(P)(z-y)}  =O(\tau^{-1}).
\eey
Therefore,
$$
\{\psi(z),\psi(y)\}=0,\qquad\qquad\qquad\qquad y\leq z.
$$
Using skew symmetry of the bracket and interchanging $y$ and $z$ we have
$$
\{\psi(z),\psi(y)\}=0,\qquad\qquad\qquad\qquad y\geq z.
$$
Taking the sum of these two formulas,  we obtain \ref{pp}.

Now we compute the bracket \ref{thi}. Let $Q,P\in \G_R$ and  $Q\rightarrow P_+,\; P\rightarrow P_-$ along the imaginary direction. Then  using formulas \ref{psib}, \ref{psi} and Lemma \ref{rrt} for  $f(x)\in C^{\infty}_0$ and $y \leq z$ we have: 
\bey
&\phantom{000}&\int\limits_{-\infty}^{z} dy  f(y) \{\psib(z),\psi(y)\}=\int\limits_{-\infty}^{z} dy f(y) \{-i\psib(z),i \psi(y)\}\\
&=&\lim\int\limits_{-\infty}^{z} dy f(y) \{\frac{\l(Q)}{\P(z,Q)},\l(P)\P(y,P)\}\\
&=& \lim -\frac{\l(Q)\l(P)}{\P^2(z,Q)}\int\limits_{-\infty}^{z} dy f(y) \{\P(z,Q),\P(y,P)\}\\
&=& \lim -\frac{\l(Q)\l(P)}{\P^2(z,Q)}\{\P(z,Q),\P(z,P)\}\int\limits_{-\infty}^{z} dy f(y) e^{-i\l(P)(z-y)}.
\eey
Using Theorem \ref{ahb}, 
\bey
\{\P(z,Q),\P(z,P)\}= 2\times \frac{\(\P(z,Q)-\P(z,P)\)^2}{\l(Q)-\l(P)}.
\eey
Since $\P$ has a pole at $P_+$ and a zero at $P_-$,  we have asymptotically
\bey
-\frac{\l(Q)\l(P)}{\P^2(z,Q)}&&\{\P(z,Q),\P(z,P)\}\sim\\
 &&\sim -\frac{\l(Q)\l(P)}{\P^2(z,Q)}\times 2 \times \frac{\P^2(z,Q)}{\l(Q)-\l(P)}.
\eey
Let $P=\ea Q$ and  $\l(Q)=i\tau,\; \tau \rightarrow +\infty$; then 
\bey
-\frac{\l(Q)\l(P)}{\P^2(z,Q)}&&\{\P(z,Q),\P(z,P)\}\sim -\frac{2\tau^2}{i\tau +i\tau}.
\eey
Using   steepest descent we have 
\bey
\therefore&=& \lim_{\tau \rightarrow +\infty} -\frac{2\tau^2}{i\tau +i\tau}\int\limits_{-\infty}^{z} dy f(y)  e^{-\tau(z-y)}=if(z).
\eey
Therefore,
\beq\label{alo}
\{\psib(z),\psi(y)\}=i \d(z-y),\qquad\qquad\qquad\qquad y\leq z.
\eeq
Since the bracket is real, by Lemma   \ref{pbreal} we have 
$$
\{\psi(z),\psib(y)\}=-i \d(z-y),\qquad\qquad\qquad\qquad y\leq z.
$$
By the skew symmetry of the bracket and  interchanging $z$ and $y$, 
\beq\label{aloo}
\{\psib(z),\psi(y)\}=i \d(z-y),\qquad\qquad\qquad\qquad z\leq y.
\eeq
Taking the sum of \ref{alo} and \ref{aloo}, we obtain  \ref{thi}.
\qed

\vskip .2in
\noindent
Department of Mathematics
\newline
Michigan State University
\newline
East Lansing, MI 48824
\newline
USA
\vskip 0.3in
\noindent
vaninsky@math.msu.edu

\end{document}

%% file: def23.tex
\newcommand{\lb}{\linebreak}
\newcommand{\noi}{\noindent}

\newcommand{\C}{{\mathbb C}}
\newcommand{\RB}{{\mathbb R}}

\newcommand{\CP}{{\mathbb C}{\mathbb  P}}

\newcommand\Dg{\mathfrak D}

\newcommand{\HH}{{\mathcal H}}
\newcommand{\MM}{{\mathcal M}}

\def\P{\mathcal X}
\def\[{\left[}
\def\]{\right]}
\def\({\left(}
\def\){\right)}

\def\l{\lambda}
\def\lb{\overline{\l}}

\def\m{\mu}
\def\lt{\frac{\lambda}{2}}
\def\psib{\overline{\psi}}

\def\d{\delta}
\def\a{\alpha}
\def\ea{\epsilon_a}

\def\b{\beta}

\def\G{\Gamma}

\def\12{{1\over 2}}

\newcommand{\e}{{\boldsymbol e}}

\newcommand{\f}{{\boldsymbol f}}
\newcommand{\pp}{{\boldsymbol p}}
\newcommand{\uu}{{\boldsymbol u}}

\newcommand{\LLL}{{\boldsymbol L}}

\newcommand{\gb}{{\boldsymbol g}}

\newcommand{\eeq}{\end{equation}}
\newcommand{\beq}{\begin{equation}}
\newcommand{\bay}{\begin{eqnarray}}
\newcommand{\ey}{\end{eqnarray}}
\newcommand{\bey}{\begin{eqnarray*}}
\newcommand{\eey}{\end{eqnarray*}}

\newtheorem{thm}{\hspace{\parindent}Theorem}[section]

\newtheorem{lem}[thm]{\hspace{\parindent}Lemma}

\newtheorem{exa}[thm]{Example}